# Intelligent Architectures for Intelligent Computing Systems

Onur Mutlu
ETH Zurich
omutlu@gmail.com

## Abstract

Computing is bottlenecked by data. Large amounts of application data overwhelm storage capability, communication capability, and computation capability of the modern machines we design today. As a result, many key applications' performance, efficiency and scalability are bottlenecked by data movement. In this invited special session talk, we describe three major shortcomings of modern architectures in terms of 1) dealing with data, 2) taking advantage of the vast amounts of data, and 3) exploiting different semantic properties of application data. We argue that an intelligent architecture should be designed to handle data well. We show that handling data well requires designing architectures based on three key principles: 1) data-centric, 2) data-driven, 3) data-aware. We give several examples for how to exploit each of these principles to design a much more efficient and high performance computing system. We especially discuss recent research that aims to fundamentally reduce memory latency and energy, and practically enable computation close to data, with at least two promising novel directions: 1) *processing using memory*, which exploits analog operational properties of memory chips to perform massively-parallel operations in memory, with low-cost changes, 2) *processing near memory*, which integrates sophisticated additional processing capability in memory controllers, the logic layer of 3D-stacked memory technologies, or memory chips to enable high memory bandwidth and low memory latency to near-memory logic. We discuss how to enable adoption of such fundamentally more intelligent architectures, which we believe are key to efficiency, performance, and sustainability. We conclude with some guiding principles for future computing architecture and system designs. This accompanying short paper provides a summary of the invited talk and points the reader to further work that may be beneficial to examine.

## I. Introduction

Existing computing systems process increasingly large amounts of data. Data is key for many modern (and likely even more future) workloads and systems. Important workloads (e.g., machine learning, artificial intelligence, genome analysis, graph analytics, databases, video analytics, online collaboration), whether they execute on cloud servers or mobile systems are all data intensive; they require efficient processing of large amounts of data. Today, we can generate more data than we can process, as exemplified by the rapid increase in the data obtained in astronomy observations and genome sequencing [1].

Unfortunately, the way they are designed, modern computers are not efficient at dealing with large amounts of data: large amounts of application data greatly overwhelm the storage capability, the communication capability, and the computation capability of the modern machines we design today. As such, data becomes a large performance and energy bottleneck, and it greatly impacts system robustness and security as well. As a prime example, we provide evidence that the potential for new genome sequencing technologies, such as nanopore sequencing [2, 113], is greatly limited by how fast and how efficiently we can process the huge amounts of genomic data the underlying technology can provide us with [3, 83, 113, 119, 143]. A similar observation can also be made for video analytics [163, 7] and machine learning [198-199, 7].

The processor-centric design paradigm (and the resulting processor-centric execution model) of modern computing systems is one prime cause of why data overwhelms modern machines [4, 5, 120]. With this paradigm, there is a clear dichotomy between processing and memory/storage: data has to be brought from storage and memory units to computation units (e.g., general-purpose processors or special-purpose accelerators), which are far away from the memory/storage units, before any processing can be done on the data. The dichotomy exists at the macro-scale (e.g., across the internet) as well as the micro-scale (e.g., within a single compute node, or even within a single CPU processing core). This processor-memory dichotomy leads to large amounts of data movement across the entire computing system, degrading performance and expending large amounts of energy. For example, a recent work [7] shows that more than 60% of the entire mobile system energy is spent on data movement across the memory hierarchy when executing four major commonly-used consumer workloads, including machine learning inference, video processing and playback, and web browsing. Similarly, due to the current processor-centric design paradigm, a large fraction of the system resources is dedicated to units that store and move data (i.e., to serve the computation units), and actual computation units constitute only ~5% of an entire processing node [8] – yet, even then, data access is still a major bottleneck due to the large latency and energy costs of accessing large amounts of data.

## II. Fundamental Principles

Our starting axiom for an intelligent architecture is that it should handle (i.e., store, access, and process) data well. But, what does it mean for an architecture to handle data well? We posit (and later demonstrate with examples) that the answer lies in satisfying three major desirable properties (or principles): 1) data-centric, 2) data-driven, and 3) data-aware.

First, the system should ensure that data does not overwhelm its components. Doing so requires effort in intelligent algorithms, intelligent architectures and intelligent whole system designs that are co-optimized cross-layer (i.e., optimizations spanning across algorithms-architectures-devices), in a manner that puts data and its processing at the center of the design, minimizing data movement and maximizing the efficiency with which data is handled, i.e., stored, accessed, and processed (e.g., as exemplified in [4-38, 120]). We call this first principle *data-centric architectures*.

Second, an intelligent architecture takes advantage of the vast amounts of data and metadata that flow through the system, to continuously improve its decision making, by bettering both its policies and mechanisms based on online learning and self-optimization. In other words, the architecture should make data-driven, self-optimizing decisions in its components (e.g., as exemplified in [39-51, 121]). We call this second principle *data-driven architectures*.

Third, an intelligent architecture understands and exploits various properties of each piece of data so that it can improve and adapt its algorithms, mechanisms, and policies based on the characteristics of data. In other words, the architecture should make *data-characteristics-aware* decisions in its components and across the entire system (e.g., as exemplified in [52-58, 107, 116, 11, 149]). We call this third principle *data-aware architectures*.

## III. EXISTING COMPUTING ARCHITECTURES

Based on our qualitative and quantitative analyses, we find that existing computing architectures greatly fall short of handling data well. In particular they violate all of the three major desirable principles. We analyze each briefly next.

First, modern architectures are poor at dealing with data: they are designed to mainly store and move data, as opposed to actually compute on the data. Most system resources serve the processor (and accelerators) without being capable of processing data. As such, existing architectures are *processor-centric* as opposed to *data-centric*: they place the most value in the processor (not data) and everything else in the system is viewed as secondary serving the processor. We believe this is the wrong mindset and approach in designing a balanced system that handles data well: such a system should be data centric: i.e., data should be the prime thing that is valued and everything else in the system should be designed to 1) minimize data movement by enabling computation capability at and close to where data resides and 2) maximize the value and efficiency of processing data by enabling low-latency and low-energy access to as well as low-energy and low-cost storage of vast amounts of data. Doing so would eliminate the huge data access bottleneck of processor-centric systems, thereby improving performance, reducing energy consumption, alleviating off-chip bandwidth requirements (and hence area and cost), likely reducing system and hardware design complexity, as well as opening up new opportunities for improving system security and reliability by handling data more locally in or near where it resides.

Second, modern architectures are poor at taking advantage of vast amounts of data (and metadata) available to them during online operation and over time. They are designed to make simple decisions based on fixed policies, ignoring massive amounts of easily-available data. This is because existing architectural policies make *human-driven* decisions as opposed to *data-driven* decisions, and humans, by nature, do not seem capable of designing policies and heuristics that consider hundreds, if not thousands, of different state attributes that may be useful to examine in a control policy that makes dynamic decisions. It is instructive to notice that a modern memory controller, for example, keeps executing *exactly* the same *fixed* policy for scheduling or power management (e.g., FR-FCFS [59, 60], PAR-BS [61] or some other heuristic-based policy [62-73, 117-118, 122-133]), during the *entire lifetime* of a system (for many many years!), regardless of the positive or negative impact of the decisions resulting from the policy at any given point of time on the system. The same is true for a modern prefetch controller, a cache controller, a network controller, and for many other hardware controllers in a system (e.g., [150-162, 200-214]). Each controller sees a vast amount of data and makes a vast number of decisions even in the timeframe of a single millisecond (let alone years), yet it is incapable of learning from that data and changing its policy to another dynamically-determined better policy because the policy it follows is rigid and hardcoded by a human. This is clearly not intelligent: for example, as humans, we have the capability to learn from the past and adapt our actions accordingly to not repeat the same mistakes as in the past or to choose the best policy/actions that we believe will provide the highest benefits in the future. Enabling similar intelligence and far-sightedness in controller and system policies in an architecture is necessary for obtaining good performance and efficiency (as well as better reliability, security and perhaps other metrics) under a variety of system conditions and workloads.

Third, modern architectures are poor at knowing and exploiting different properties of application and system data. They are designed to treat all data as the same (except for a small set of specialized hints that provide some opportunity to optimize based on data characteristics in a limited manner that is very specific to the particular optimization). As such, the decisions existing architectures make are *component-aware* decisions as opposed to *data-aware* decisions: a component's (e.g., a cache's or a memory controller's) structural and performance characteristics dominate the policies designed to control that component and the accessed/manipulated data's characteristics are rarely conveyed to the policies or even known. If the characteristics of the data to be accessed or manipulated were known, the decisions taken could be very different: for example, if we knew the relative compressibility of different types of data, e.g., different data types or different objects [55, 74-81, 135-138], different components in the entire system could be designed in a manner that adaptively scales their capability to match the compressibility of different data elements, in order to maximize both performance and efficiency. Modifying the architecture and its interface to become richer and more expressive, and to include rich and accurate information on various properties of data that is to be processed, is therefore critical to customizing the architecture to the characteristics of the data and, thus, enabling intelligent adaptation of system policies to data characteristics.

## IV. INTELLIGENT COMPUTING ARCHITECTURES

A major chunk of our invited talk describes in detail the characteristics of an intelligent computing architecture, by concrete examples and their empirical evaluation. This short paper does not go into detail, but provides a brief overview with references to other works that exemplify such architectures. Multiple detailed versions of this talk can be found online [82, 139-142]. We also refer the reader to recent detailed survey and overview papers we have written on the topic [120, 4].

### Data-Centric

A data-centric architecture has at least four major characteristics. First, it enables processing capability in or near where data resides (i.e., in or near memory structures), as described in detail in [4-6, 8, 38, 120] and exemplified by [7-12, 14, 19, 20, 24, 27, 30, 34, 84, 108-113, 144-147]. Second, it provides low-latency and low-energy access to data, as exemplified by [11-13, 15-18, 21, 23, 31-33, 84-86]. Third, it enables low-cost data storage and processing (i.e., high capacity memory at low cost, via techniques like new memory technologies, hybrid memory systems and/or compressed memory systems), as exemplified by [22, 87-96, 74, 76, 78, 107, 116]. Fourth, it provides mechanisms for intelligent data management (with intelligent controllers handling robustness, security, cost, etc.), as described in detail in [97-103, 116, 120] and exemplified by, e.g., [104-106, 116, 120, 179-190].

Our talk provides significant detail on providing processing capability in or near where data resides, focusing on processing in memory (PIM). There is a pressing need for enabling PIM in modern systems due to 1) *a bottom-up push*, i.e., circuit- and device-level memory technology scaling issues requiring intelligent main memory controllers to solve low-level scaling and reliability challenges, such as RowHammer [104-106, 99, 102], data retention [21, 167-170, 97, 191-193], energy consumption [171-172, 127, 132], enabling scalable emerging technologies [22, 87-93, 172-174] and 2) *a top-down pull*, i.e., systems and applications requiring near-data processing capability with minimal data movement to reduce the data access bottleneck and its large negative effect on performance [154-155, 164-165], energy [7, 166], and sustainability.

There are at least two new approaches to enabling processing-in-memory in modern systems. The first approach, *processing using memory (PUM)*, exploits the existing memory architecture and the operational principles of the memory circuitry to enable operations inside memory structures with minimal changes. PUM makes use of intrinsic properties and operational principles of the memory cells and cell arrays, by inducing interactions between cells such that the cells and/or cell arrays can perform useful computation. PUM architectures enable a wide range of different functions, such as data copy/initialization, bitwise operations, and simple arithmetic operations. We focus on how to minimally and practically change DRAM chips to perform fast and energy-efficient bulk data copy and initialization [84, 12, 147, 175] as well as bulk bitwise operations [6, 10, 109, 175]. Similar approaches are also applicable to SRAM, MRAM, RRAM and other NVM technologies [176-178].

The second approach, *processing near memory (PNM)*, involves adding or integrating computation units (e.g., accelerators, simple processing cores, reconfigurable logic) close to or inside the memory. Computation units can be placed in the logic layer of 3D-stacked memories, in the memory controller, or even inside memory chips. Recent advances in silicon interposers (in-package wires that connect directly to the through-silicon vias in a 3D-stacked chip) also allow for separate logic chips to be placed in the same die package as a 3D-stacked memory while still taking advantage of the TSV bandwidth.

Both PUM and PNM approaches can greatly accelerate real applications, including database systems, graph analytics, machine learning, genome analysis, GPU workloads, pointer-chasing-intensive workloads, data analytics, climate modeling, etc. Recent results show up to approximately two orders of magnitude improvement in energy and performance over conventional processor-centric systems. More functionality can be potentially integrated into a memory chip using PNM than using PUM, but both approaches can be combined to get even higher benefit from PIM. For both approaches, we describe and tackle relevant cross-layer research, design, and practical adoption challenges in devices, architecture, systems, and programming models in our talk. Our recent PIM overview work comprehensively analyzes modern PIM systems and issues [120, 4].

*Data-Driven*

A data-driven architecture enables the machine itself to learn the best policies for managing itself and executing programs. Controllers in such an architecture, when needed, are data-driven autonomous agents that automatically learn far-sighted policies. A prime example of such a controller is the *reinforcement learning based self-optimizing memory controllers* [39]. Such controllers can not only improve performance and efficiency under a wide variety of conditions and workloads but also reduce the hardware and system designer's burden in designing sophisticated controllers [39]. We believe an intelligent architecture will consist of a collection of such intelligent controllers that perform automatic data-driven online policy learning, including learning of how to best coordinate with each other to make decisions that benefit the overall system. Such machines learn the best policies over time and thus become better as they learn, adapting, evolving, and executing far-sighted policies. To enable such a machine, we need to revisit the design of all controllers (e.g., caching, prefetching, storage, memory, interconnect) and turn them into data-driven agents.

*Data-Aware*

A data-aware architecture understands what it can do with and to each piece of data (and associated computations on data), and uses this information about data characteristics to maximize system efficiency and performance. In other words, it customizes itself (i.e., its policies and mechanisms) to the characteristics of the data and computations it is dealing with. Such an architecture requires knowledge of various characteristics of different data elements and structures as well as computations. Many semantic or other characteristics of data (e.g., compressibility, approximability, sparsity, criticality, access and security semantics, locality, latency vs. bandwidth sensitivity, privacy requirements, data types, error vulnerability) are invisible or unknown to modern hardware and thus need to be communicated or discovered. We believe efficient and expressive software/hardware interfaces and resulting cross-layer mechanisms, as exemplified by X-Mem (Expressive Memory) [52, 53] and the Virtual Block Interface [56] as well as other works [54, 55, 57, 58, 107, 116, 11], are promising and critically-needed approaches to creating general-purpose data-aware architectures.

ACKNOWLEDGMENTS

An earlier version of this talk was delivered as a plenary keynote talk at the VLSI-DAT/TSA conferences [142], with an accompanying paper [148], which this paper is an extension of. The very first version of this talk was delivered as a keynote talk at the SRC-Mubadala-Khalifa Forum on The Future of Artificial Intelligence Hardware Systems in April 2019. We thank all of the members of the SAFARI Research Group, and our collaborators at Carnegie Mellon, ETH Zurich, and other universities, who have contributed to the various works we describe in this paper. Thanks also goes to our research group's industrial sponsors over the past ten years, especially ASML, Google, Huawei, Intel, Microsoft, NVIDIA, Samsung, Seagate, SRC, and VMware, who have supported various pieces of research that are described in this paper and the associated talk.